\documentclass[aps,pre,%
twocolumn,%
groupedaddress,floats,amssymb,amsmath,amsfonts,floatfix]{revtex4}

\usepackage{graphicx}

\begin{document}


\title{Anisotropic Energy Distribution in Three-Dimensional Vibrofluidized Granular Systems}


\author{Peter E. Krouskop}
\affiliation{Department of Chemistry and Biochemistry, Duquesne University, Pittsburgh, PA 15282-1530, USA}
\author{Julian Talbot}
\email[Corresponding author. ]{talbot@duq.edu}
\affiliation{Department of Chemistry and Biochemistry, Duquesne University, Pittsburgh, PA 15282-1530, USA}


\date{\today}

\begin{abstract}
We examine the energy distribution in a three-dimensional model granular system consisting of $N$ inelastic hard spheres contained in an open cylinder of radius $R$ under the influence of gravity.  Energy is supplied to the system by a vibrating base.  We introduce spatially resolved, partial particle-particle ``dissipations'' $\mathcal{D}^{\|}_{pp}(\mathbf{r})$ and $\mathcal{D}^{\perp}_{pp}(\mathbf{r})$ for directions parallel and perpendicular to the energy input, respectively.  Energy balances show that the total (integrated) ``dissipation'' $D^{\|}_{pp} < 0$ while $D^{\perp}_{pp} > 0$. The energy supplied to the perpendicular directions is dissipated by particle-wall collisions.  We further define $\beta = -D^{\perp}_{pp}/D^{\|}_{pp}$, which in the steady state represents the fraction of the power supplied by the vibrating base that is dissipated at the wall.  We examine the dependence of $\beta$ on the number of particles, the velocity of the vibrating base, the particle-particle restitution coefficient, and the particle-wall restitution coefficient.  We also explore the influence of the system parameters on the spatially dependent partial dissipations.  We observe that, unlike $\mathcal{D}^{\|}_{pp}(\mathbf{r})$, $\mathcal{D}^{\perp}_{pp}(\mathbf{r})$ changes sign.
\end{abstract}


\maketitle

\section{Introduction \label{intro}}

In an isolated granular system of inelastic particles the energy-dissipating collisions rapidly lead to a total cessation of motion.  If an energy source is present, such as a vibrating wall, a fluidized steady state can be maintained.  The steady state is characterized by the balance of the energy input with the energy dissipated through particle-particle and particle-boundary collisions. Although there is a superficial resemblance, these non-equilibrium steady state granular systems have radically different properties from their equilibrium counterparts in systems of elastic particles.  Examples include non-Maxwellian velocity distributions \cite{GT1996,EP1997,VE1998,LCDKG1999,BP2000,FM,EB}, a lack of energy equipartition between rotational and translational degrees of freedom in a system of inelastic rough spheres \cite{ML1998}, and a lack of equipartition of kinetic energy between the components of a granular mixture \cite{MP1999, BDS1999, GD1999, WP2002, MP2002, CH2002, MP2002a, PMP2002, BT}. 

Moreover, as has been observed in several studies, the breakdown of equipartition in granular systems extends to the translational degrees of freedom.  In an experimental study of a three-dimensional vibrofluidized system Wildman et al.\ \cite{WHP2001} showed that the granular temperature in the direction parallel to the energy injection is larger than in the perpendicular directions.  Morgado and Mucciolo \cite{MM}, using a variant of the Direct Simulation Monte Carlo method, demonstrated a similar effect in a two-dimensional vibrated granular system.  Prevost et al.\ \cite{PEU2002} showed that velocity correlations in vibrated monolayers are different in the longitudinal and transverse directions.  Barrat et al.\ \cite{BTF2001} studied a vertically shaken three-dimensional system and noted that the injected energy is transferred to the perpendicular directions through particle-particle collisions. This results in apparent restitution coefficients that are greater than unity for some collisions, and lead Barrat et al.\ to propose the Random Restitution Coefficient model in which the restitution coefficient of each collision is randomly sampled from a given distribution \cite{BTF2001, BT2002b, BTc}.

Various theories of vibrated granular media have been proposed \cite{LHB1994,WHJ1995,K1998,H1998,ML1998a, SK}.  Warr et al.\ \cite{WHJ1995} proposed expressions for the particle-particle dissipation and the power supplied to obtain a scaling relationship for the average granular temperature. Their experimental results were not consistent with the theory, however.  Kumaran \cite{K1998} sought to improve the theory by using a Maxwellian velocity distribution. This had the effect of modifying the prefactors but not the exponents of the scaling relations.  McNamara and Luding \cite{ML1998a} reviewed and extended the theories and compared their predictions to the results of event driven simulations. The comparison confirmed that the Kumaran scaling relationship applies only in dilute systems.  Sunthar and Kumaran \cite{SK} subsequently extended Kumaran's analysis to dense systems.  More recently, Brey et al.\ \cite{BR-MM2001} used a hydrodynamic description of an open vibrated system in two and three dimensions.  They showed that the temperature profile exhibits a minimum as a function of height, and also obtained an accurate description of the energy dissipated.  

In this paper we explore the anisotropy of the energy consumed or supplied by particle-particle collisions in a three-dimensional, vibrofluidized system of inelastic, hard spheres. More specifically, we resolve the total energy supply or consumption into directions parallel and perpendicular to the energy source. Energy balances show that collisions consume energy in the parallel direction and supply it to the perpendicular directions. We examine the dependence of these quantities on the particle-particle and particle-wall restitution coefficients as well as the velocity of the vibrating base. We also examine the dependence of the \emph{local} energy consumption/supply on position and the above system parameters.   

\section{Simulation \label{simulatn}}
\subsection{Simulation algorithms}
\begin{figure}
\includegraphics[height=4in,keepaspectratio=true]{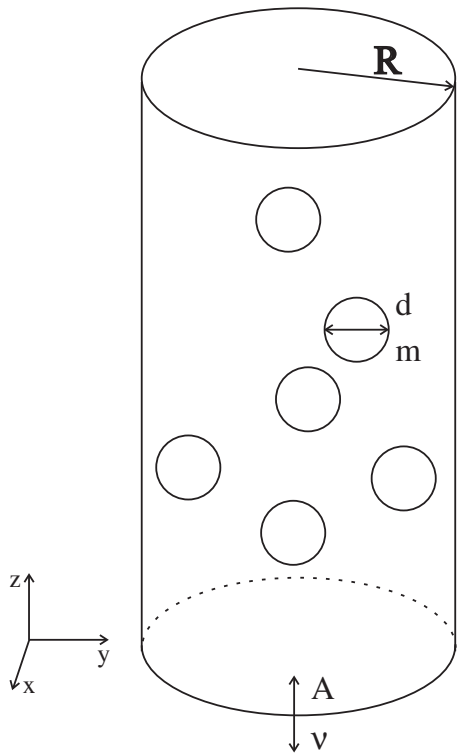}
\caption{The model system of $N$ particles of mass $m$ and diameter $d$ in an infinitely tall cylinder of radius $R$.  The base is vibrated with an amplitude $A$ and a frequency $\nu$.  \label{fig:sys}}
\end{figure}

We use a three-dimensional, event-driven simulation of inelastic, hard spheres to model granular particles in a cylinder in the presence of gravity.  Identical spheres with a diameter $d$ and mass $m$ are placed in an infinitely tall cylinder with a radius $R$ (see Figure \ref{fig:sys}). The simulation generates a sequence of particle-particle, particle-wall and particle-base collisions. The base vibrates with a symmetric, saw-tooth waveform with amplitude $A$ and frequency $\nu$.  This form was chosen for simplicity, and changing to another symmetric form, e.g. a sine wave, is not expected to have much influence on the steady state properties of the system \cite{MB1997}.  The velocity of the base for this type of waveform is $v_b = 4 A \nu$.

The simulations were initialized by randomly placing particles in the cylinder and assigning velocities selected from a Maxwell distribution.  The particles were then allowed to evolve from the initial configuration to a steady state before data was collected for the average values presented in this paper.

The particle-particle collisions modeled in the computer simulation conserve momentum, but dissipate energy.  Although several models have been proposed for inelastic collisions \cite{BTF2001, LM}, we use the simplest approach, i.e. a constant restitution coefficient.  This choice has produced results that are in good agreement with experiment \cite{MP1999, BDS1999, WP2002, WHJ1995, TV2002, KT2003}.  The restitution coefficient ranges between 0 (a completely inelastic collision) and 1 (an elastic collision).  The post-collision velocities of colliding particles $i$ and $j$ are related to the pre-collision velocities ($\mathbf{v_i}$ and $\mathbf{v_j}$) by the following equations: 
\begin{subequations}
\label{eqn:vel_pp}
\begin{equation}
\label{eqn:vel_pp1}
\mathbf{v'_i} = \mathbf{v_i} - \frac{1+c}{2} \left[ \left(\mathbf{v_i}-\mathbf{v_j}\right) \cdot \mathbf{\hat{n}}) \right] \mathbf{\hat{n}}
\end{equation}
\begin{equation}
\label{eqn:vel_pp2}
\mathbf{v'_j} = \mathbf{v_j} + \frac{1+c}{2} \left[ \left(\mathbf{v_i}-\mathbf{v_j}\right) \cdot \mathbf{\hat{n}}) \right] \mathbf{\hat{n}},
\end{equation}
\end{subequations}
where $c$ is the restitution coefficient and the unit vector $\mathbf{\hat{n}}$ points from the center of particle $i$ to the center of particle $j$.  Similarly, the equation for a particle-wall collision with a restitution coefficient of $c_w$ is 
\begin{equation}
\label{eqn:vel_pw}
\mathbf{v'_i} = \mathbf{v_i} - \left(1 + c_w\right) \left(\mathbf{v_i} \cdot \mathbf{\hat{r}}\right) \mathbf{\hat{r}},
\end{equation}
where $\mathbf{\hat{r}}$ is a unit vector that points from the center of particle $i$ to the point of contact with the wall.  Finally, the equation for a particle-base collision with a restitution coefficient of $c_b$ is 
\begin{equation}
\label{eqn:vel_pb}
\mathbf{v'_i} = \mathbf{v_i} -\left(1 + c_b\right) \left[\left(\mathbf{v_i} - \mathbf{v_b}\right) \cdot \mathbf{\hat{r}} \right] \mathbf{\hat{r}},
\end{equation}
where $\mathbf{v_b}$ is the velocity of the base.

Equations \ref{eqn:vel_pp} and \ref{eqn:vel_pw} result in energy changes of 
\begin{equation}
\label{eqn:de_pp}
\Delta E_{ij} = - \frac{m \left(1-c^2 \right)}{4} \left[ \left(\mathbf{v_i} - \mathbf{v_j}\right) \cdot \mathbf{\hat{n}} \right]^2
\end{equation}
and 
\begin{equation}
\label{eqn:de_pw}
\Delta E^{wall}_i = -\frac{m \left(1-c_w^2 \right)}{2} \left(\mathbf{v_i} \cdot \mathbf{\hat{r}}\right)^2
\end{equation}
for particle-particle and particle-wall collisions, respectively.  These energy changes are always negative for values of $c$ and $c_w$ less than 1.  The energy change associated with a particle colliding with the base is 
\begin{eqnarray}
\label{eqn:de_pb}
\Delta E^{base}_i = \frac{m}{2} \left[ \left(1+c_b \right)^2 \left[ \left(\mathbf{v_i} - \mathbf{v_b}\right) \cdot \mathbf{\hat{r}} \right]^2 \right] \nonumber\\
- m \left(1+c_b \right) \left(\mathbf{v_i} \cdot \mathbf{\hat{r}}\right) \left[ \left(\mathbf{v_i} - \mathbf{v_b}\right) \cdot \mathbf{\hat{r}} \right],
\end{eqnarray}
which can be either positive or negative.  Most collisions with the base impart energy to the particle, although if the collision occurs as the base is descending energy is lost.

\subsection{System Properties}

The simulated trajectories of the particles are used to calculate macroscopic properties of the system such as the packing faction ($\eta$), the granular temperature ($T_{\alpha}$) and the energy dissipation.  

To facilitate the comparison of the simulation results with different experimental systems we use dimensionless parameters.  The characteristic length and mass are taken as the particle diameter $d$ and mass $m$, respectively.  A natural characteristic time is $\sqrt{d/g}$ where $g$ is the gravitational constant.  All numerical results presented in this paper are dimensionless and can be converted to real units using the conversion factors presented in Table \ref{tab:param}.

\begin{table}
\caption{Relations for converting between dimensionless (quantities with a star) and real (quantities without a star) quantities. \label{tab:param}}
\begin{tabular}{|c|c|}
\hline
Conversion Equation&Parameter\\
\hline
$R^*=R/d$&Radius\\
$H^*=H/d$&Height\\
$A^*=A/d$&Amplitude\\
$\nu^*=\nu \left(d/g\right)^{1/2}$&Frequency\\
$v^*=v /\left(g d\right)^{1/2}$&Velocity\\
$T^*=T /\left(m g d \right)$&Temperature\\
$E^*=E /\left(m g d \right)$&Energy\\
\hline
\end{tabular}
\end{table}

The packing fraction is defined as the ratio of the volume occupied by particles to the sample volume:
\begin{equation}
\eta = \frac{\left< v_p \right>}{V} = \frac{\left< n \right> \pi d^3}{6 V},
\end{equation}
where $\left< v_p \right>$ is the average volume occupied by particles within a sample volume $V$, and $\left< n \right>$ is the average number of particles in this volume.  

\begin{figure*}
\includegraphics[height=4in,keepaspectratio=true]{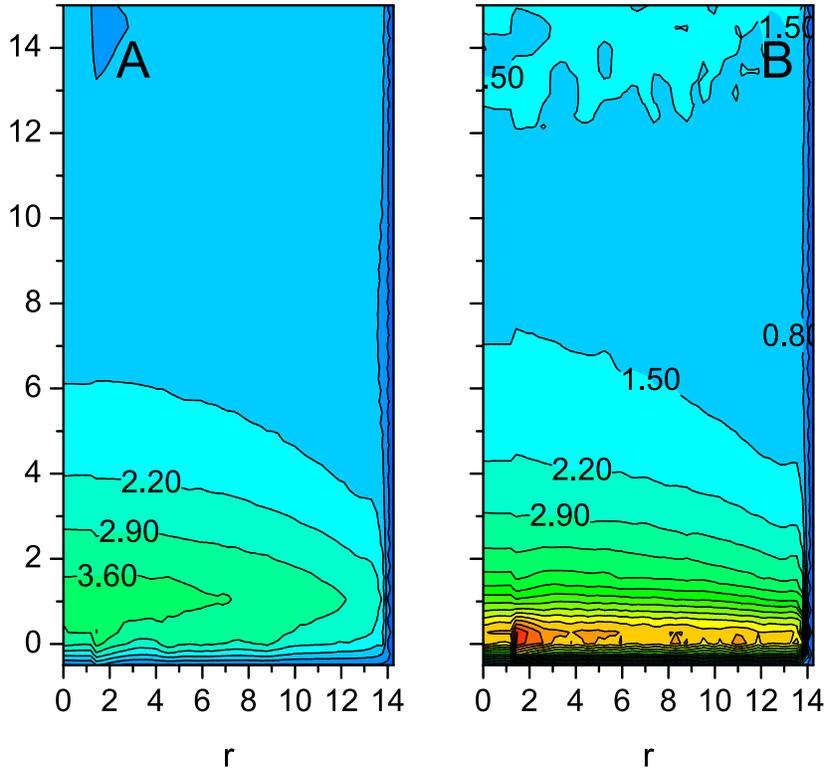}
\caption{The granular temperature ($T_{\alpha}$) as a function of height and radial position in (A) the x and y directions and (B) the z direction.  The contours designate a change in $T_{\alpha}$ of $0.7$.  Simulation conditions: $N=2100$, $R=14.5$, $A=0.348$, $\nu=1.13$, $c=0.91$, and $c_w = 0.68$.  \label{fig:temp_2d}}
\end{figure*}

The granular temperature in the $\alpha$ direction is calculated as
\begin{equation}
\label{eqn:temp}
T_{\alpha} = m \left<  v_{\alpha}^2 \right>,
\end{equation}
where $\left< v_{\alpha}^2 \right>$ is the mean squared velocity in the $\alpha$ direction.    As shown in previous work, these are not equal and vary with position \cite{WP2002, WHP2001, MM, KT2003}.  The directional granular temperatures for a system consisting of 2100 particles is shown in Figure \ref{fig:temp_2d}.

To study the dissipation in the system we start with the overall energy balance:
\begin{equation}
\label{eqn:total_e_bal}
\sum_{n_p} \Delta E_{ij} + \sum_{n_w} \Delta E^{wall}_i + \sum_{n_b} \Delta E^{base}_i + \Delta \Phi + \Delta K = 0
\end{equation}
where $n_p$ is the number of particle-particle collisions, $n_w$ is the number  of particle-wall collisions, and $n_b$ is the number of particle-base collisions.    $\Delta \Phi$ and $\Delta K$ are the differences in the potential and kinetic energies, respectively, of the system at the start and the end of the simulation:
\begin{equation}
\label{eqn:delta_phi}
\Delta \Phi = \sum_{i=1}^N m g \left(z_{i,final} - z_{i,initial} \right),
\end{equation}
and
\begin{equation}
\label{eqn:delta_k}
\Delta K = \sum_{i=1}^N \frac{1}{2} m \left(v^2_{i,final} - v^2_{i,initial} \right).
\end{equation} 
where e.g. $z_{i,initial}$ and $v_{i,initial}$ are the initial values of the height and velocity of particle $i$, respectively. The final values are designated with the subscript \emph{final}.

$\Delta \Phi$ and $\Delta K$ are bounded and fluctuate around steady state values.  The other sums in the total energy balance, however, are unbounded as they increase linearly with the total number of collisions, $n = n_p + n_w + n_b$. Therefore, for a sufficiently large number of collisions corresponding to an elapsed time $t$ Equation \ref{eqn:total_e_bal} simplifies to: 
\begin{equation}
\label{eqn:total_e_lim}
D_{pp} + D_{pw} + P_b = 0,
\end{equation}
where 
\begin{equation}
D_{pp}=\frac{1}{t}\sum_{n_p}\Delta E_{ij}
\end{equation}
is the dissipation due to particle-particle collisions,
\begin{equation}
D_{pw}=\frac{1}{t}\sum_{n_p}\Delta E^{wall}_i
\end{equation}
is the dissipation due to particle-wall collisions and
\begin{equation}
P_b=\frac{1}{t}\sum_{n_p}\Delta E^{base}_i 
\end{equation} is the power supplied by the vibrating base.

Motivated by the concept of directional granular temperatures for the system (e.g. Equation \ref{eqn:temp}) we separate the energy changes into the three degrees of freedom for translation.  For example, the energy change in the z-direction due to a particle-particle collision is 
\begin{eqnarray}
\label{eqn:de_pp_dir}
\Delta E^{\|}_{ij} = \frac{m}{4} \left(1+c \right)^2 \left(\mathbf{v_r} \cdot \mathbf{\hat{n}} \right)^2 \left(\mathbf{\hat{n}} \cdot \mathbf{\hat{z}} \right)^2 \nonumber\\
- \frac{m}{2} \left(1+c \right) \left(\mathbf{v_r} \cdot \mathbf{\hat{n}} \right) \left(\mathbf{\hat{n}} \cdot \mathbf{\hat{z}} \right) \left(\mathbf{v_r} \cdot \mathbf{\hat{z}} \right),
\end{eqnarray}
where $\mathbf{v_r} = \mathbf{v_i} - \mathbf{v_j}$ is the relative velocity of the two particles and $\mathbf{\hat{z}}$ is a unit vector pointing in the z-direction.  A similar equation defines the energy change in the x- and y-directions, $\Delta E^{\perp}_{ij}$.  The two energy changes are related by
\begin{equation}
\label{eqn:de_relation}
\Delta E_{ij}=\Delta E^{\perp}_{ij}+\Delta E^{\|}_{ij} \le 0
\end{equation}
We may then write two additional energy balances -- one for the direction parallel to the energy input by the base (i.e. the z-direction), and another for the perpendicular directions (i.e. the x- and y-directions):
\begin{equation}
\label{eqn:par_e_bal}
\sum_{n_p} \Delta E^{\|}_{ij} + \sum_{n_b} \Delta E^{\|}_i + \Delta \Phi + \Delta K^{\|} = 0
\end{equation}
and 
\begin{equation}
\label{eqn:perp_e_bal}
\sum_{n_p} \Delta E^{\perp}_{ij} + \sum_{n_w} \Delta E^{\perp}_i + \Delta K^{\perp} = 0,
\end{equation}
where the superscripts $\|$ and $\perp$ denote the parallel and perpendicular directions, respectively.  Since the particles are taken as smooth $\Delta E^{\|}_i = 0$ for collisions with the wall and $\Delta E^{\perp}_i = 0$ for base collisions.  Gravity acts in the parallel direction, hence the potential energy change, $\Delta \Phi$, appears only in Equation \ref{eqn:par_e_bal}. In the limit of a large number of collisions, Equations \ref{eqn:par_e_bal} and \ref{eqn:perp_e_bal} become
\begin{equation}
\label{eqn:par_e_lim}
D^{\|}_{pp} + P_b = 0
\end{equation}
and
\begin{equation}
\label{eqn:perp_e_lim}
D_{pp}^{\perp} + D_{pw} = 0.
\end{equation}
Since energy is never gained at the wall, $D_{pw} \le 0$, and on average input at the base, $P_b>0$, it follows from Equations \ref{eqn:par_e_lim} and \ref{eqn:perp_e_lim} that
\begin{equation}
\label{eqn:par_ppsum}
D^{\|}_{pp} < 0
\end{equation}
and
\begin{equation}
\label{eqn:perp_ppsum}
D^{\perp}_{pp} \ge 0.
\end{equation}
Like $D_{pp}$, $D^{\|}_{pp}$ is always negative (for inelastic collisions). $D^{\perp}_{pp}$, however, is positive, i.e. the collisions supply energy to the perpendicular directions. $D^{\perp}_{pp}$ does not, therefore, represent dissipation of energy. The equality in Equation \ref{eqn:perp_ppsum} applies only when the walls of the cylinder are elastic and $\Delta E^{\perp}_i = 0$.  Also since the total energy change due to a collision is either zero for an elastic collision or negative for $c<1$
\begin{equation}
\label{eqn:par_perp_ppsum}
D_{pp} = D^{\|}_{pp} + D^{\perp}_{pp} \le 0.
\end{equation}
Thus, the amount of energy lost in the parallel direction is always greater than the energy gained in the perpendicular direction.

Since we are interested in the amount of energy that is transferred to the perpendicular direction we find it convenient to define the \emph{fractional energy transfer} 
\begin{equation}
\label{eqn:beta}
\beta = - \frac{D^{\perp}_{pp}}{D^{\|}_{pp}}
\end{equation}
which by Equation \ref{eqn:par_perp_ppsum} is bounded between zero and one.  
In the steady state it follows from Equations \ref{eqn:par_e_lim} and \ref{eqn:perp_e_lim} that $\beta = -D_{pw}/P_b$. Thus, in the steady state $\beta$ can be interpreted as the fraction of the power supplied by the vibrating base that is dissipated by particle-wall collisions. Similarly, the
fraction of the power dissipated by particle-particle collisions is 
$-D_{pp}/P_b=1-\beta$. 

We also investigate the spatial dependence of the particle-particle dissipation by introducing
\begin{equation}
\mathcal{D}^{\perp}_{pp}(r,z)=\frac{1}{t}\sum_{n_p}\Delta E^{\perp}_{ij}
\end{equation}
where the summation is over collisions that occur in the volume element at $(r,z)$. A similar equation defines $\mathcal{D}^{\|}_{pp}(r,z)$. The total system dissipation in the perpendicular direction is 
\begin{equation}
D^{\perp}_{pp}=2 \pi \int_0^{\infty} \int_0^{R} \mathcal{D}^{\perp}_{pp}(r,z) r \, \mathrm{d}r \, \mathrm{d}z
\end{equation}
A similar expression gives the relation between $D^{\|}_{pp}$ and $\mathcal{D}^{\|}_{pp}(r,z)$.  In addition to $\mathcal{D}^{\perp}_{pp}(r,z)$ we also examine the reduced distributions
\begin{equation}
\mathcal{D}^{\perp}_{pp}(z)=2 \pi \int_0^{R}\mathcal{D}^{\perp}_{pp}(r,z) r \, \mathrm{d}r
\end{equation}
as well as $\mathcal{D}^{\|}_{pp}(z)$, which is defined similarly.

\section{Results and Discussion}
We used the simulation to examine the dependence of  $\beta$ on the base velocity, the particle-particle restitution coefficient, and the particle-wall restitution coefficient for systems of varying particle number $N$.  Additionally, we investigated the dependence of $\mathcal{D}^{\|}_{pp}(r,z)$ and $\mathcal{D}^{\perp}_{pp}(r,z)$ on the packing fraction and temperature distribution in the system.  Finally, we studied the trends in the radially averaged functions $\mathcal{D}^{\|}_{pp}(z)$ and $\mathcal{D}^{\perp}_{pp}(z)$ as the number of particles, the velocity of the base, the particle-particle restitution coefficient, and the particle-wall restitution coefficient were varied.  Constant values of $R=14.5, A=0.348$ and $c_b=0.88$ were used in all simulations presented in this paper. 

\subsection{Fractional energy transfer}
\begin{figure}
\includegraphics[width=3.375in,keepaspectratio=true]{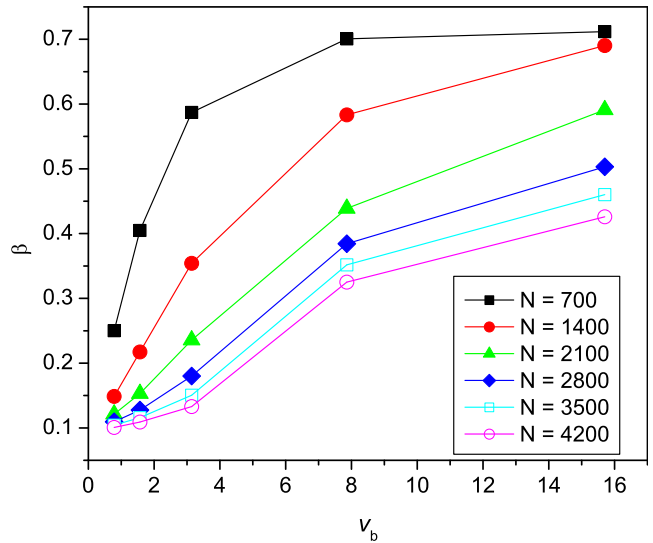}
\caption{Values of $\beta$ for systems with 700 to 4200 particles and base velocities ranging from 0.786 to 15.7.  Simulation conditions were $R=14.5$, $A=0.348$, $c=0.91$, and $c_w = 0.68$.  \label{fig:sys_v_deratio}}
\end{figure}

Figure \ref{fig:sys_v_deratio} shows the effect of the base velocity on the transfer of energy between the parallel and perpendicular directions.  Systems consisting of 700 to 4200 particles were simulated with base velocities of $v_b = 0.786, 1.57, 3.14, 7.86, 15.7$.  $\beta$ increases with increasing base velocity at constant $N$ and decreases with increasing $N$ at constant $v_b$. The former trend is due to the decrease in density as the granular temperature increases, which leads to more particle-wall collisions.  Thus, more of the energy input by the base is being dissipated at the wall.  Increasing $N$ at constant $v_b$ increases the number of particle-particle collisions. The increase in the particle-particle collision rate reduces the fraction of the input power dissipated at the wall. 

\begin{figure}
\includegraphics[width=3.375in,keepaspectratio=true]{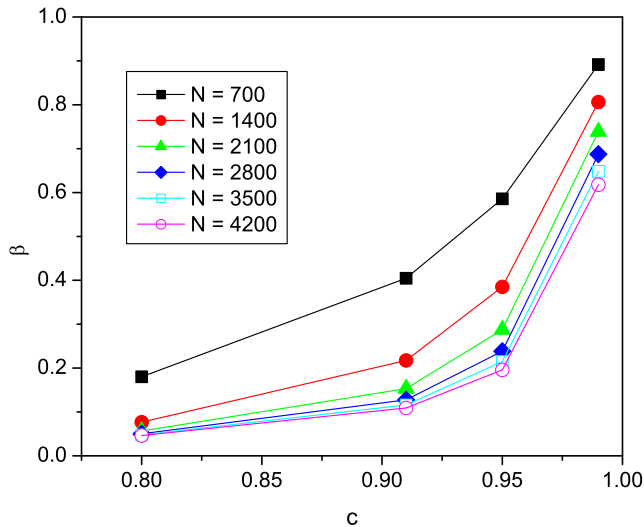}
\caption{Values of $\beta$ for systems with 700 to 4200 particles and particle-particle restitution coefficients ranging from 0.8 to 0.99.  Simulation conditions were $R=14.5$, $A=0.348$, $\nu=1.13$, and $c_w = 0.68$.  \label{fig:sys_c_deratio}}
\end{figure}

\begin{figure}
\includegraphics[width=3.375in,keepaspectratio=true]{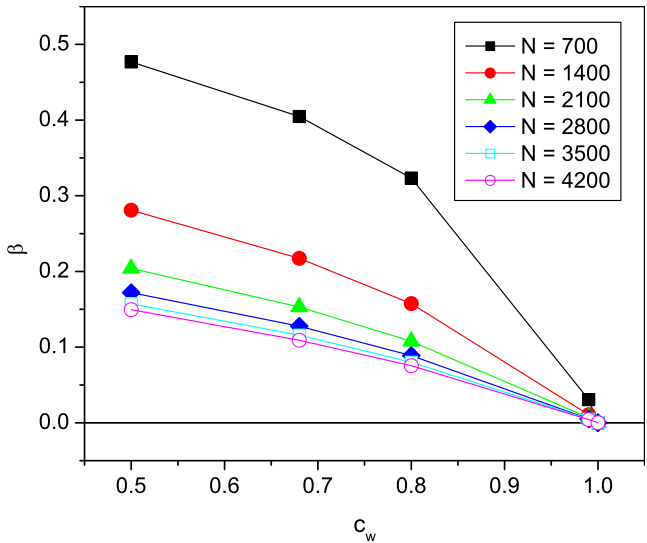}
\caption{Values of $\beta$ for systems with 700 to 4200 particles and particle-wall restitution coefficients ranging from 0.5 to 1.00.  Simulation conditions were $R=14.5$, $A=0.348$, $\nu=1.13$, and $c=0.91$.  \label{fig:sys_cw_deratio}}
\end{figure}

It is interesting to note that $\beta$ appears to approach an asymptotic limit as $v_b$ increases, at least for 700 and 1400 particle systems. We expect that the limit is dependent on the values of the particle-particle restitution coefficient $c$ and the particle-wall restitution coefficient $c_w$.  The data in Figures \ref{fig:sys_c_deratio} and \ref{fig:sys_cw_deratio} support this hypothesis.

The fractional energy transfer was also calculated for systems of 700 to 4200 particles with particle-particle restitution coefficients ranging from $0.8 \le c \le 0.99$.  Figure \ref{fig:sys_c_deratio} shows the values of $\beta$ as a function of $c$ for each system size.  As the particle-particle restitution coefficient approaches unity the value of $\beta$ increases.  When $c=1$, according to Equation \ref{eqn:par_perp_ppsum},  $\beta=1$ as all power must be dissipated at the wall.  Figure \ref{fig:sys_c_deratio} also shows that $\beta$ decreases as the number of particles increases, as seen previously.

We have devised phenomenological explanations for these trends.  As the particle-particle restitution coefficient increases, the granular temperature increases, and the particle bed expands leading to fewer particle-particle collisions and more particle-wall collisions as explained above.  It appears that as the number of particles increases, $\beta$ becomes independent of $N$. As the system becomes denser, a greater fraction of the power is dissipated by particle-particle collisions. Eventually the power supplied is insufficient to sustain a fluidized system. The particles simply collapse and move as a whole on the vibrating base. This last situation cannot be studied directly with the current version of the simulation.

The effect of varying the particle-wall restitution coefficient at constant $N$ and $v_b$ is shown in Figure \ref{fig:sys_cw_deratio}.  As the restitution coefficient goes to one, $\beta$ goes to zero.  The elastic wall dissipates no energy, so the dissipation $D^{\perp}_{pp}$ goes to zero according to Equation \ref{eqn:perp_e_lim}.  As $c_w$ decreases the wall is able to dissipate more energy.  As in the other graphs, an increase in the number of particles is seen to decrease the value of $\beta$.  

\subsection{Local dissipation}
We investigated the dissipation as a function of position in systems containing 2100 particles (roughly three layers of particles resting on the base). We determined the packing fraction and the dissipation in the parallel and perpendicular directions as functions of height and radial position.
 
\begin{figure*}
\includegraphics[height=4in,keepaspectratio=true]{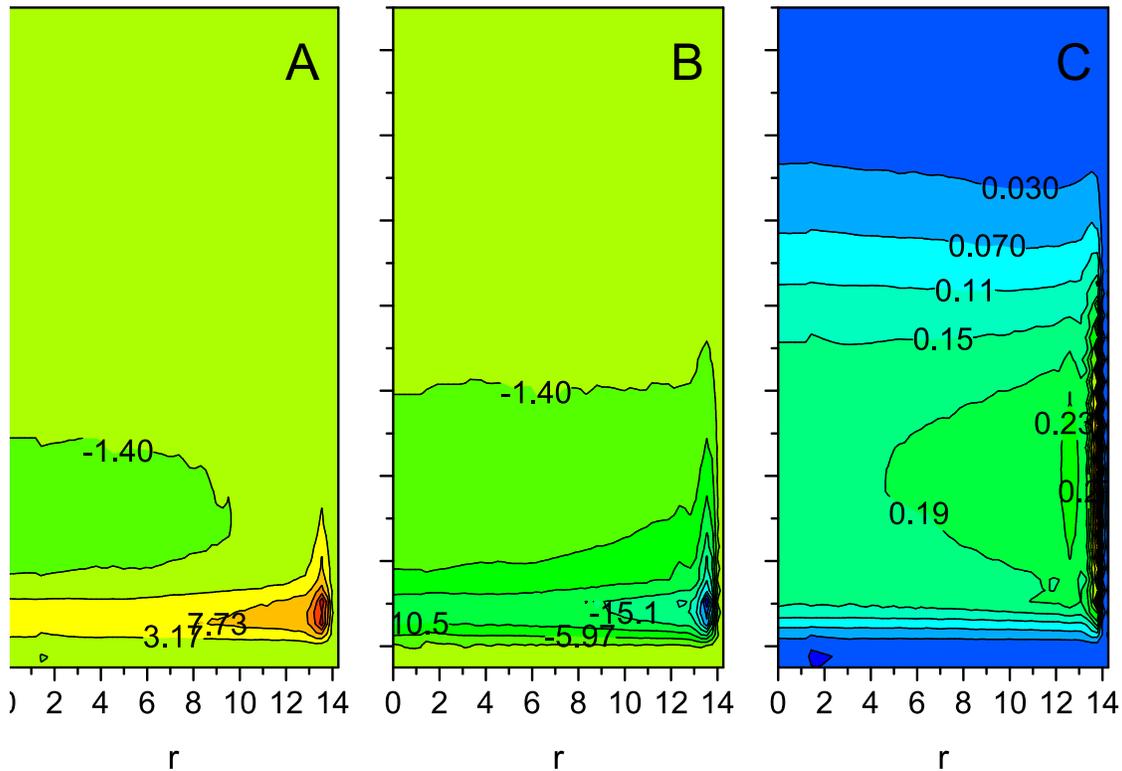}
\caption{Contour plots of (A) $\mathcal{D}^{\perp}_{pp}(r,z)$, (B) $\mathcal{D}^{\|}_{pp}(r,z)$, and (C) packing fraction as a function of height and radial position.  The contours designate a change of $4.57$ in graphs A and B.  Simulation conditions are the same as those for Figure \ref{fig:temp_2d}.  \label{fig:de_dens_2d}}
\end{figure*}

Figure \ref{fig:de_dens_2d}A shows a contour plot of $\mathcal{D}^{\perp}_{pp}(r,z)$ and Figure \ref{fig:de_dens_2d}B shows a contour plot of $\mathcal{D}^{\|}_{pp}(r,z)$.  While the latter is everywhere negative, energy is both dissipated and generated in the perpendicular direction, depending on the position. The dissipation shows radial and height dependencies for both the perpendicular and parallel directions similar to those observed for the packing fraction shown in Figure \ref{fig:de_dens_2d}C.  The extrema in the dissipation profiles, however, occur closer to the base than the maximum in the packing fraction.  The maximum in $\mathcal{D}^{\perp}_{pp}(r,z)$ occurs at the same location as the minimum in $\mathcal{D}^{\|}_{pp}(r,z)$.  It should be noted that the minimum in the parallel direction is larger in magnitude than the maximum in the perpendicular direction as required by Equation \ref{eqn:de_relation}.  The variations in $\mathcal{D}^{\perp}_{pp}(r,z)$ and $\mathcal{D}^{\|}_{pp}(r,z)$ as a function of height are different, as shown in Figure \ref{fig:de_dens_2d}.  Specifically, $\mathcal{D}^{\perp}_{pp}(r,z)$ is positive at small heights with a negative minimum in the middle while $\mathcal{D}^{\|}_{pp}(r,z)$ is negative over the entire region, i.e., energy is never gained in the parallel direction from particle-particle collisions.  It is interesting to note that the trends observed for $\mathcal{D}^{\perp}_{pp}(r,z)$ and $\mathcal{D}^{\|}_{pp}(r,z)$ appear to be the inverse of those seen in the temperature profiles presented in Figure \ref{fig:temp_2d}.

\subsection{Influence of the number of particles on local dissipation}
We varied the number of particles in the system from 700 to 4200, corresponding to approximately 1 to 6 layers of particles at rest.  The base velocity was kept constant at $v_b=1.57$.

\begin{figure}
\includegraphics[width=3.375in,keepaspectratio=true]{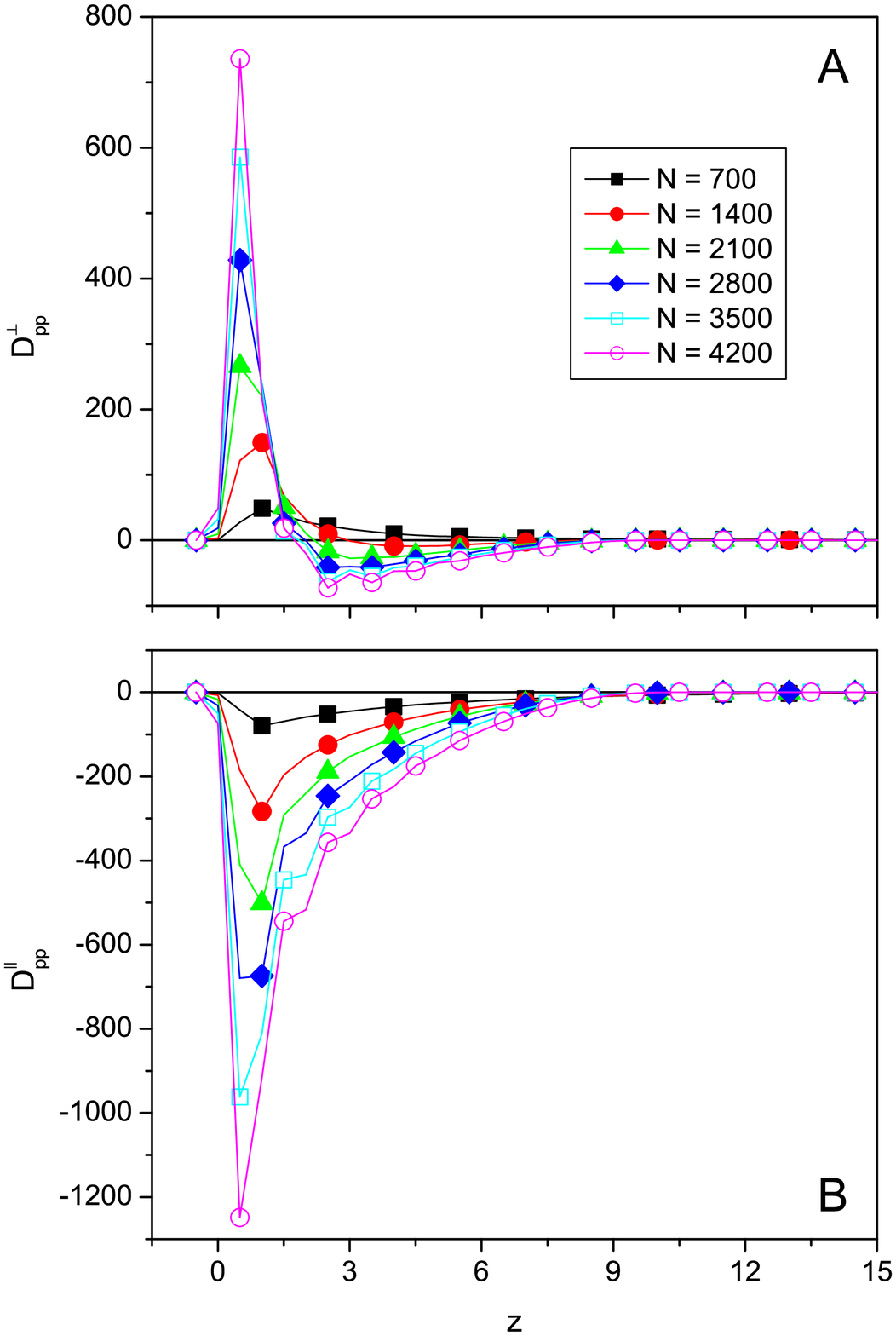}
\caption{The dissipation in the (A) perpendicular and (B) parallel directions as a function of height for values of $N$ ranging from 700 to 4200 particles (1 to 6 layers of particles).  Other simulation conditions are $R=14.5$, $A=0.348$, $\nu=1.13$, $c=0.91$, and $c_w = 0.68$.  \label{fig:n_de}}
\end{figure}

The values of $\mathcal{D}^{\perp}_{pp}(z)$ and $\mathcal{D}^{\|}_{pp}(z)$ for each system are shown in Figure \ref{fig:n_de}.  The dissipation in the perpendicular direction exhibits a positive maximum for all the systems studied.  In addition, the systems with 1400 or more particles display a negative minimum.  The 700-particle system does not show a minimum in $\mathcal{D}^{\perp}_{pp}(z)$ indicating that the perpendicular energy dissipation in this system occurs through particle-wall collisions.  Figure \ref{fig:n_de}A shows that as the number of particles increases, the maximum in $\mathcal{D}^{\perp}_{pp}(z)$ increases and the minimum decreases. There is also a small shift in the height at which the maximum is observed.  The shift in the positions of the extrema to smaller heights is due to the successive cooling and compression of the bed as the number of particles is increased.

The dissipation in the parallel direction is always negative and exhibits a minimum for all the systems as seen in Figure \ref{fig:n_de}B.  The position of the minimum decreases from around $z=1.0$ to $z=0.5$ as the number of particles increases.  The heights of these minima coincide with those observed for the maxima in the perpendicular direction.  It should also be noted that the maximum loss in $\mathcal{D}^{\|}_{pp}(z)$ is greater than the maximum gain in $\mathcal{D}^{\perp}_{pp}(z)$, as expected for inelastic particles.

\subsection{Influence of the base velocity on local dissipation}
\begin{figure}
\includegraphics[width=3.375in,keepaspectratio=true]{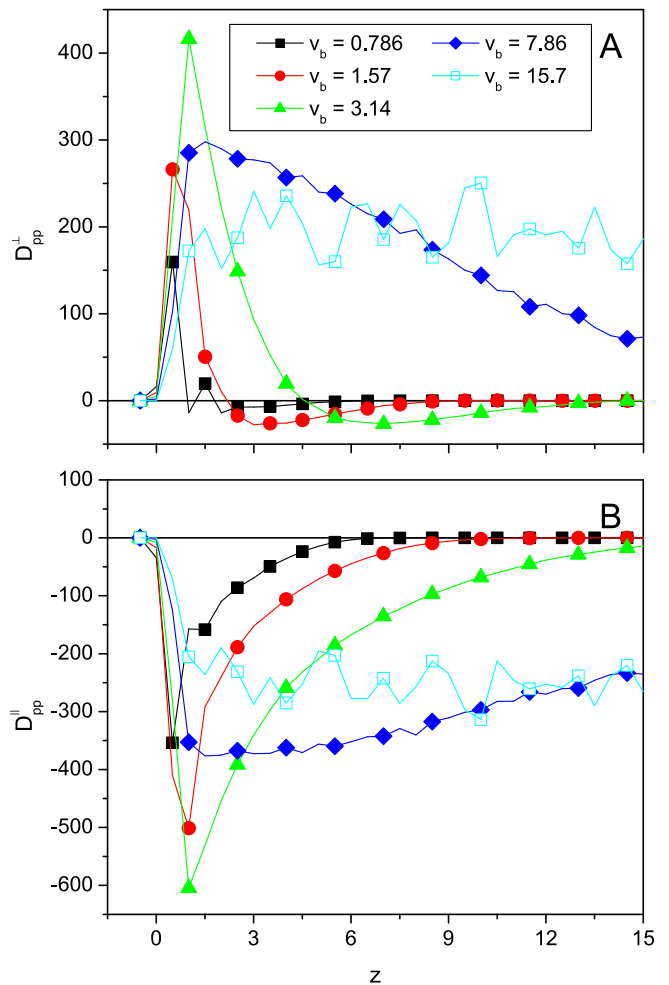}
\caption{The dissipation in the (A) perpendicular and (B) parallel directions for systems in which the velocity of the base ($v_b$)was varied from 0.786 to 15.7.  The simulation conditions were $N=2100$, $R=14.5$, $A=0.348$, $\nu=0.565, 1.13, 2.26, 5.65, 11.3$, $c=0.91$, and $c_w = 0.68$.  \label{fig:v_de}}
\end{figure}

We also examined the effect of varying the base velocity on the dissipation.  Systems consisting of 700 to 4200 particles were simulated with base velocities of $0.786 \le v_b \le 15.7$.  Figure \ref{fig:v_de} displays the dissipation in the perpendicular and parallel directions for a system with 2100 particles at each base velocity.  An increase in the base velocity from 0.786 to 3.14 results in an increase in the positive maximum of $\mathcal{D}^{\perp}_{pp}(z)$.  There is a corresponding decrease in the minimum observed in $\mathcal{D}^{\|}_{pp}(z)$.  Furthermore, the maximum in the perpendicular direction shifts to a larger height as the velocity is increased.  The shifts in the height are related to the expansion of the bed.  The velocities of $v_b = 7.86$ and 15.7 exhibit a decrease in the magnitude of the extrema for the perpendicular and parallel directions.  The decrease does not translate to a decrease in energy transfer, however, since the curves broaden noticeably for the larger velocities.  The energy transfer is no longer confined to a small region of heights, but now extends over a larger region.  It is also interesting that the values of $\mathcal{D}^{\perp}_{pp}(z)$ are always positive for the two largest velocities.

\subsection{Influence of the particle-particle restitution coefficient on local dissipation}
\begin{figure}
\includegraphics[width=3.375in,keepaspectratio=true]{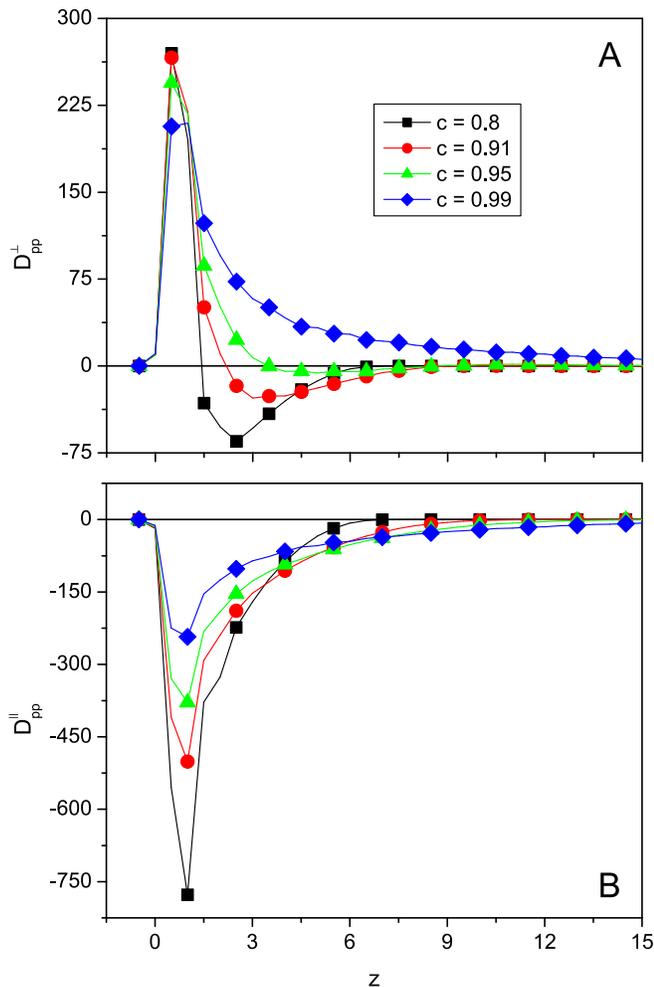}
\caption{The dissipation in the (A) perpendicular and (B) parallel directions as a function of height for systems in which the particle-particle restitution coefficient $c$ was varied from 0.8 to 0.99.  The simulation conditions were $N=2100$, $R=14.5$, $A=0.348$, $\nu=1.13$,and $c_w = 0.68$. \label{fig:c_rate}}
\end{figure}

We also collected the values of $\mathcal{D}^{\|}_{pp}(z)$ and $\mathcal{D}^{\perp}_{pp}(z)$ for systems in which the particle-particle restitution coefficient ($c$) was changed (Figure \ref{fig:c_rate}).  We varied $c$ from 0.8 to 0.99 for a system with 2100 particles at a constant base velocity of $v_b=1.57$.  The dissipation in the perpendicular direction (Figure \ref{fig:c_rate}A) reaches the highest maximum and lowest minimum for the system with the restitution coefficient of $c=0.8$.  These extrema decrease in magnitude as the restitution coefficient increases, and for $c = 0.99$ there is no observable minimum in $\mathcal{D}^{\perp}_{pp}(z)$.  Unlike the previous studies of the velocity, these curves do not broaden.  The peak does not shift and there is only a small decrease in height and increase in breadth.  The dissipation in the parallel direction (Figure \ref{fig:c_rate}B) is always negative, as seen previously.  The minima become shallower as the restitution coefficient increases.  It should be noted that for the nearly elastic system ($c = 0.99$) the magnitude of the minimum in $\mathcal{D}^{\|}_{pp}(z)$ is nearly equal to the magnitude of the maximum in $\mathcal{D}^{\perp}_{pp}(z)$ as would be expected for an elastic system.

\subsection{Influence of the particle-wall restitution coefficient on local dissipation}
The energy balance, Equation \ref{eqn:perp_e_lim} implies that the energy dissipated at the wall will influence $\mathcal{D}^{\perp}_{pp}(z)$.  We investigated this relationship by simulating systems with varying values of the particle-wall restitution coefficient.  The results for a system consisting of 2100 particles and particle-wall restitution coefficients of $c_w = 0.5, 0.68, 0.8, 0.99$ are presented below.

\begin{figure}
\includegraphics[width=3.375in,keepaspectratio=true]{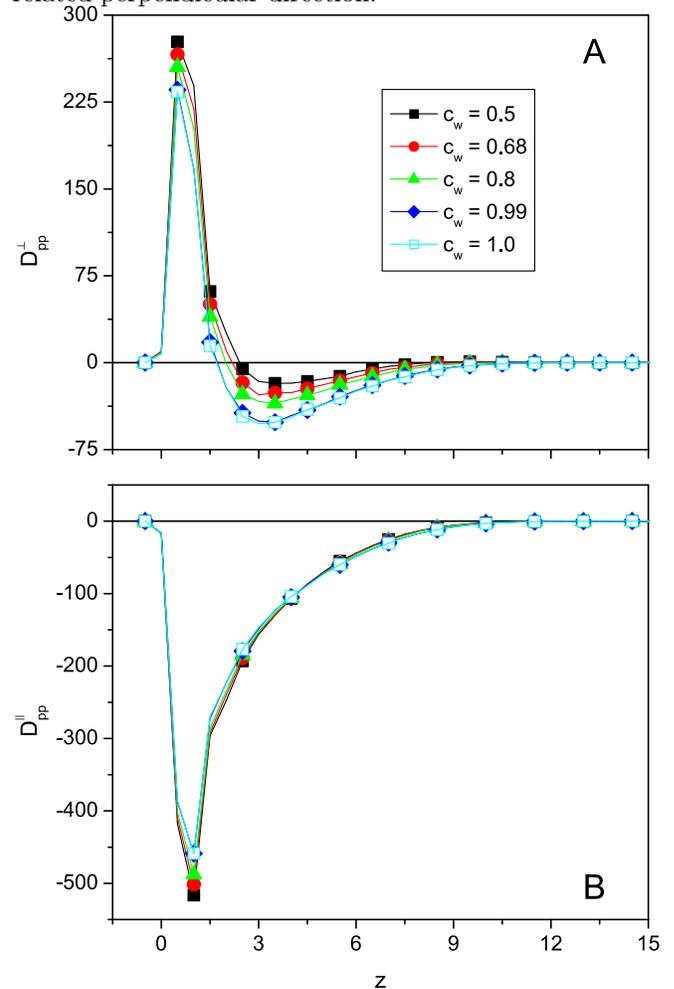}
\caption{The dissipation in the (A) perpendicular (B) parallel directions as a function of height for systems in which the particle-wall restitution coefficient $c_w$ was varied.  The simulation conditions were $N=2100$, $R=14.5$, $A=0.348$, $\nu=1.13$, and $c=0.91$. \label{fig:cw_rate}}
\end{figure}

The values of $\mathcal{D}^{\perp}_{pp}(z)$ and $\mathcal{D}^{\|}_{pp}(z)$ as a function of height are presented in Figure \ref{fig:cw_rate}.  The dissipation in the perpendicular direction (Figure \ref{fig:cw_rate}A) does not vary greatly, but there is a noticeable trend.  The maximum observed at small heights becomes less positive and the minimum becomes more negative as $c_w$ approaches unity.  This indicates that more energy is lost through particle-particle collisions as the particle-wall restitution coefficient approaches unity.  The minima in $\mathcal{D}^{\|}_{pp}(z)$, as seen in Figure \ref{fig:cw_rate}B, indicate that there is less energy loss in the parallel direction as the particle-wall restitution coefficient approaches unity.

\section{Conclusions \label{concl}}
In this paper we have examined in detail the energy transfer between the direction parallel to the energy source and the perpendicular directions in a driven granular system.  Energy balances show that the energy changes resulting from particle-particle collisions in these directions $D^{\|}_{pp}$ and $D^{\perp}_{pp}$ are always negative and positive, respectively, for inelastic systems.  We introduced the fractional energy transfer, $\beta = - D^{\perp}_{pp}/D^{\|}_{pp}$, which in the steady state equals the fraction of energy input at the base that is dissipated at the wall.  We further examined the quantitative dependence of these quantities on the number of particles, the base velocity, the particle-particle restitution coefficient, and the particle-wall restitution coefficient.  We then examined the local dissipations, $\mathcal{D}^{\|}_{pp}(r,z)$ and $\mathcal{D}^{\perp}_{pp}(r,z)$.  While the former is negative throughout the system, the latter changes sign.

When an energy source is present in an experimental system, it is always anisotropic to some degree, i.e. energy is supplied preferentially in some direction(s).  Therefore the differences in the parallel and perpendicular directions observed here must always be present in real systems.  Systems with lower symmetry than the one studied here, e.g. a rectangular prism, are expected, in addition, to exhibit differences in each non-symmetry-related perpendicular direction.

\begin{acknowledgments}
We would like to thank the National Science Foundation (CHE-9814236) for financial support.
\end{acknowledgments}

\bibliography{granmat}

\begin{thebibliography}{10}
\expandafter\ifx\csname url\endcsname\relax
  \def\url#1{\texttt{#1}}\fi
\expandafter\ifx\csname urlprefix\endcsname\relax\def\urlprefix{URL }\fi

\bibitem{GT1996}
I.~Goldhirsch, M.-L. Tan, Phys. Fluids 8 (1996) 1752.

\bibitem{EP1997}
S.~E. Esipov, T.~P{\"{o}}schel, J. Stat. Phys. 86 (1997) 1385.

\bibitem{VE1998}
T.~P.~C. van Noije, M.~H. Ernst, Granular Matter 1 (1998) 57.

\bibitem{LCDKG1999}
W.~Losert, D.~G.~W. Cooper, J.~Delour, A.~Kudrolli, J.~P. Gollub, Velocity
  statistics in excited granular media, Chaos 9 (1999) 682.

\bibitem{BP2000}
N.~Brilliantov, T.~P{\"o}schel, Deviation form maxwell distribution in granular
  gasses with constant restitution coefficient, Phys. Rev. E 61 (2000) 2809.

\bibitem{FM}
K.~Feitosa, N.~Menon, Breakdown of energy equipartition in a 2d binary vibrated
  granular gas (2001).

\bibitem{EB}
M.~Ernst, R.~Brito, Scaling solutions of inelastic boltzmann equations with
  over-populated high energy tails (2002).

\bibitem{ML1998}
S.~McNamara, S.~Luding, Energy nonequipartition in systems of inelastic, rough
  spheres, Phys. Rev. E 58 (1998) 2247.

\bibitem{MP1999}
P.~A. Martin, j.~Piasecki, Mean-field model of free-cooling inelastic mixtures,
  Europhys. Lett. 46 (1999) 613.

\bibitem{BDS1999}
J.~J. Brey, J.~W. Dufty, A.~Santos, J. Stat. Phys. 97 (1999) 281.

\bibitem{GD1999}
V.~Garzo, J.~Dufty, Homogeneous cooling state for a granular mixture, Phys.
  Rev. E 60 (1999) 5706.

\bibitem{WP2002}
R.~D. Wildman, D.~J. Parker, Coexistence of two granular temperatures in binary
  vibrofluidized beds, Phys. Rev. Lett. 88 (2002) 064301.

\bibitem{MP2002}
U.~M.~B. Marconi, A.~Puglisi, Mean-field model of free-cooling inelastic
  mixtures, Phys. Rev. E 65 (2002) 051305.

\bibitem{CH2002}
R.~Clelland, C.~M. Hrenya, Simulations of a binary mixture of inelastic grains
  in rapid shear flow, Phys. Rev. E 65 (2002) 031301.

\bibitem{MP2002a}
U.~M.~B. Marconi, A.~Puglisi, Steady-state properties of a mean-filed model of
  driven inelastic mixtures, Phys. Rev. E 66 (2002) 011301.

\bibitem{PMP2002}
R.~Pagnani, U.~M.~B. Marconi, A.~Puglisi, Driven low density mixtures, Phys.
  Rev. E 66 (2002) 051304.

\bibitem{BT}
A.~Barrat, E.~Trizac, Lack of energy equipartition in homogeneous heated binary
  granular mixtures (2002).

\bibitem{WHP2001}
R.~D. Wildman, J.~M. Huntley, D.~J. Parker, Convection in highly fluidized
  three-dimensional granular beds, Phys. Rev. Lett. 86 (2001) 3304.

\bibitem{MM}
W.~Morgado, E.~Mucciolo, Numerical simulation of vibrated granular gasses under
  realistic boundary conditions (2002).

\bibitem{BTF2001}
A.~Barrat, E.~Trizac, J.~N. Fuchs, Heated granular fluids: the random
  restitution coefficient approach, Eur. Phys. J. E 5 (2001) 161.

\bibitem{BT2002b}
A.~Barrat, E.~Trizac, Molecular dynamics simulations of vibrated granular
  gases, Phys. Rev. E 66 (2002) 051303.

\bibitem{BTc}
A.~Barrat, E.~Trizac, Random inelaticity and velocity fluctuations in a driven
  granular gas (2003).

\bibitem{LHB1994}
S.~Luding, H.~Herrmann, A.~Blumen, Simulations of two-dimensional arrays of
  beads under external vibrations: Scaling behavior, Phys. Rev. E 50 (1994)
  3100.

\bibitem{WHJ1995}
S.~Warr, J.~M. Huntley, G.~Jacques, Fluidization of a two-dimensional pranular
  system: Experimental study and scaling behavior, Phys. Rev. E 52 (1995) 5583.

\bibitem{K1998}
V.~Kumaran, Temperature of a granular material ``fluidized'' by external
  vibrations, Phys. Rev. E 57 (1998) 5660.

\bibitem{H1998}
J.~M. Huntley, Scaling laws for a two-dimensional vibro-fluidized granular
  material, Phys. Rev. E 58 (1998) 5168.

\bibitem{ML1998a}
S.~McNamara, S.~Luding, Energy flows in vibrated granular media, Phys. Rev. E
  58 (1998) 813.

\bibitem{SK}
P.~Sunthar, V.~Kumaran, Temperature scaling in a dense vibro-fluidised granular
  material (1999).

\bibitem{BR-MM2001}
J.~J. Brey, M.~J. Ruiz-Montero, F.~Moreno, Hydrodynamics of an open vibrated
  granular system, Phys. Rev. E 63 (2001) 061305.

\bibitem{MB1997}
S.~McNamara, J.~L. Barrat, Energy flux into a fluidized granular medium at a
  vibrating wall, Phys. Rev. E 55 (1997) 7767.

\bibitem{LM}
S.~Luding, S.~McNamara, How to handle the inelastic collapse of a dissipative
  hard-sphere gas with the {TC} model (1998).

\bibitem{TV2002}
J.~Talbot, P.~Viot, Wall-enhanced convection in vibrofluidized granular
  systems, Phys. Rev. Lett. 89 (2002) 064301.

\bibitem{KT2003}
P.~E. Krouskop, J.~Talbot, Mass and size effects in three-dimensional
  vibrofluidized granular mixtures, in press.

\end{thebibliography}

\end{document}